# Computer Simulations Reveal Motor Properties generating stable anti-parallel Microtubule interactions.

François Nédélec

Cell Biology and Biophysics Program,
European Molecular Biology Laboratory (EMBL),
Meyerhofstrasse 1, 69117 Heidelberg, Germany.
Tel: 49-6221-387-360.
Fax: 49-6221-387-242
Email: nedelec@embl-heidelberg.de

**The final formatted version of this article is available at http://www.jcb.org**

## Abstract

An aster of microtubules is a set of flexible polar filaments with dynamic plus-ends, which irradiate from a common location at which the minus ends of the filaments are found. Processive soluble oligomeric motor complexes can bind simultaneously to two microtubules, and thus exert forces between two asters. Using computer simulations, I have explored systematically the possible steady-state regimes reached by two asters under the action of various kinds of oligomeric motors. As expected, motor complexes can induce the asters to fuse, for example when the complexes consist only of minus-end directed motors, or to fully separate, when the motors are plus-end directed. More surprisingly, complexes made of two motors of opposite directionalities can also lead to anti-parallel interactions between overlapping microtubules that are stable and sustained, like those seen in mitotic spindle structures. This suggests that such hetero-complexes could have significant biological role, if they exist in the cell.

## Introduction

In eukaryotic cells, polar filaments and associated proteins play an essential role in determining intracellular order. Microtubules are often found in highly connected structures, such as the mitotic spindle in dividing cells (Wittmann et al., 2001; Karsenti and Vernos, 2001) or complex arrays in differentiated cells. These cellular assemblies can be observed and perturbed, and this has yielded invaluable insights into their modes of organization and their dynamic properties. Microtubules can be reconstituted from pure tubulin, with the optional addition of other purified cytoskeletal proteins, and these in-vitro studies have provided quantitative data about their physical properties. However, although the filaments, many associated proteins and most of the molecular motors participating in the spindle have been identified, and much is known about their individual modes of action, we still understand poorly how they participate collectively to the morphogenesis and steady-state dynamics of this structure. Questions such as: How is a balance of forces achieved? What makes it stable? are difficult to address in quantitative terms in real spindles. It seems that to examine such fundamental questions, it is desirable to study structures with intermediate levels of complexity, and to develop tools to handle them.

Computer simulations are one such tool, and they have been used before to examine the motility of polar filaments driven by immobilised motors (Bourdieu et al., 1995; Gibbons, 2001), microtubule dynamic instability (Bayley et al., 1989; Glicksman et al., 1993; Dogterom et al., 1995), actin bundle contraction (Nakazawa and Sekimoto, 1996 ; Kurse and Julicher, 2000), aster centering inside a box (Holy et al., 1997), and the formation of microtubule asters by soluble oligomeric motors (Nédélec et al. 1997, 2001; Surrey et al., 2001).

Here I describe computer simulations which calculate the evolution of a set of dynamic filaments with motor proteins. Using these stochastic simulations, I examine how two asters of dynamic microtubules nucleated by two microtubule organizing centers can reach a steady-state configuration in which microtubules overlap fully or partially. Such overlaps are essential in some spindles, to counteract the forces that pull the chromosomes apart.

## Results

To investigate theoretically whether bi-functional motors can produce a stable interaction between two microtubule asters, a simulation was built (see methods). It includes asters composed of a variable number of dynamic and flexible microtubules, and oligomeric motor-complexes which can bind up to two microtubules, ultimately producing forces between the asters.

Although the situation studied is very simple compared to that in a cell, its quantitative description already required ~27 parameters (see table 1), and the first problem was to find values for all of them. Three parameters are model-specific and do not reflect any real property (e.g. the time step dt), their values were chosen to generate sufficient numerical precision (see methods). Many of the other parameters describe known properties of the situation studied: temperature, viscosity of the fluid, microtubule rigidity and dynamic instability parameters, and were set accordingly.

What remained unspecified were the number of microtubules in each aster, and the parameters describing the motors.

There are estimates for the values that would be appropriate for conventional kinesin, and they have been successfully used to simulate the behavior of synthetic kinesin complexes in *in-vitro* experiments (Nédélec et al. 2001, Surrey et al. 2001). However, the motors involved in spindle organization have not yet been thoroughly biophysically characterized. Most of their *in-vivo* characteristics – force, speed, processivity – are still unknown. To avoid an arbitrary choice, a different approach was chosen, where these parameters were set randomly within a reasonable range (see table 1). Thus to probe the generic possibilities of the system, parameter sets were generated representing a possible number of microtubules, and combinations of motor characteristics: many or few, plus or minus directed, fast or slow, strong or weak, high or low rates.

In essence, I performed a virtual screen for possible steady states for the relative position of two asters exposed to various motor complexes. The screen was based on the automatic selection of *persisting interactions* between two asters among a collection of simulations generated automatically, with *interaction* meaning that the two asters are linked by motor complexes. Specifically, for each generated parameter set, a simulation was started from unbound motors and two randomly positioned asters, and the evolution in time was calculated for 1000 seconds of simulated time. A total of ~50,000 simulations in 1D and ~10,000 simulations in 2D were computed. From these, interactions that persist in time were selected and unstable or transient patterns discarded, by an automatic scan which retained only those simulations which had always at least one motor complex linking the two asters together, after an initial 500 s which allowed the system to equilibrate. The selected simulations fall in one of only four categories: fusion, separation or oscillations of the asters, and stable anti-parallel interactions.

Overall, 40 % of the simulations were persisting interactions. From the correlations in the corresponding parameters it is found for example that low off-rate compensates for low on-rate, or high motor concentration compensate for low on-rate (not shown). Most of these correlations simply show that the motors must have a certain *efficiency* to produce a persisting interaction: the rates and concentrations should define an equilibrium in which sufficient motors can bind two microtubules which are crossing or overlapping. This is a basic requirement which does not by itself guarantee success. Motors with inappropriate speeds, however efficient they are, never lead to persisting interactions.

Five screens have been performed (fig. 1), in which two different motors u and v were simulated. By convention, u and v will also designate the speeds of these motors, with positive values for plus-end directed motors, and negative values for minus-end directed motors. The screens differed in the way the motors were assembled into complexes: Screen 1 considered two kinds of homo-motor complexes, i.e. schematically (u-u) and (v-v). Screen 2 considered homo-motor complexes (u-u) and MAP-motor complexes (v-z), with z a motor of speed zero in the model. Screens 3a, 3b and 3c considered one kind of hetero-motor complexes (u-v). In all theses screens, the number of parameters varied (see table 1) and the number of simulations ~2000 calculated were similar.

**Screen 1:** With two sorts of homo-complexes (u-u and v-v), in all the combinations of directionality, speed, forces and rates, all persisting interactions resulted in the fusion or the full separation of the two asters. This was usually achieved within 1000s, but in rare cases of slow dynamics, the simulations had to be prolonged in time to reach fusion or full separation. A persisting interaction with a non-zero distance between the centers was not found. With only minus-end directed motors, the fused configuration is globally stable. In the presence of only homo-complexes of plus-end directed motors only, the fused situation is stable or meta-stable, but only arises if the asters are initially very close. When the asters are initially far enough apart, they separate further, under the action of motors binding anti-parallel microtubule (fig. 4).

**Screen 2:** The next screen is inspired by a proposed model of the mitotic spindle, in which the motors Eg5 and Ncd balance their forces to determine the overlap zone (Sharp et al., 1999). Eg5 is a homo-tetramer, and has been proposed to crosslink two microtubules (Kashina et al. 1996), behaving essentially like a slow (in-vitro ~ 0.03 μm/s) double plus-end directed complex. Ncd has a minus-end directed motor domain in its C-terminus (in-vitro ~ -0.25 μm/s; McDonald et al., 1990), but also an additional non-motor microtubule binding domain (Karabay and Walker, 1999). Thus Ncd can in principle also crosslink microtubules. To explore this scenario, Ncd was modeled as a hetero-complex made from a static microtubule binder, and a minus-end directed motor (v-z, with $v < 0$ and $z = 0$). Eg5 was modeled as a bi-functional plus-end directed motor (u-u, with $u > 0$). The speeds and other characteristics of these complexes were varied randomly within the usual ranges (table 1). Like in the first screen, persisting interactions only resulted in fusion or full separation of the asters.

**Screen 3a:** As shown on figure 1, when hetero-motor complexes (u-v) were simulated, the speeds of the motors simply determined which persisting interactions could appear: (1) if $u < 0$ and $v < 0$ a persisting interaction always resulted in the fusion of the asters. (2) If $u + v > 0$, persisting interaction only caused full separation of the asters, with the exception of the meta-stable situation of initially fused asters. (3) If $u \cdot v < 0$ and $u + v < 0$, persisting interactions could be of several kind. Diverse oscillatory solutions were observed (not shown, see www), provided the motors unbind at their maximum force (see methods). They were not studied further. More interesting were some persisting interactions, characterized by a complete anti-parallel microtubule overlap, in which the distance between the aster centers was never below 2 μm. By looking precisely at the motor complexes in all these simulations (www), one could see that the balance of forces was always achieved in the same way. Hence they all represented the same generic qualitative solution ($S_1$). $S_1$ was realized in about 10 % of the simulations in this screen, but the sampling was still too sparse to determine precisely the region in parameter space associated with $S_1$. However, the speeds of the motors leading to $S_1$ are found over the entire domain allowed by the two rules $u \cdot v < 0$ and $u + v < 0$ (fig. 6). Other parameters are also important in determining if the motors will produce $S_1$ or not, but the correlations found are only the ones expected to be needed to produce efficient motors.

The balance of forces defining $S_1$ can be first described in one dimension only, assuming microtubules of equal length. As pictured in figure 3, the anti-parallel overlap is always total: some microtubules always reach behind the center of the other aster. The attractive force between the asters is mostly produced by motors binding anti-parallel microtubules, and the repulsive force is produced by motors bound to parallel microtubules (fig. 4, A). Both components can be high, typically 100 pN, but cancel each other to produce an equilibrium under tension. The attractive force can be easily understood, as $u + v < 0$ is sufficient for the anti-parallel complex configurations to produce attraction (fig. 4, B). The repulsive forces are more surprising, because one should expect that motors linking two parallel microtubules exert forces that on average cancel each others out (fig. 4, C). A net force arises only if the two possible configurations (u binds $MT_1$ and v binds $MT_2$) or (v binds to $MT_1$ and u binds to $MT_2$) are not equally present. As shown on figure 3, this is exactly the case for the motors linking the parallel overlaps in $S_1$. This asymmetry is a consequence of the movement of motors which brings them to the overlapping region always in the same configuration.

The equilibrium $S_1$ is stable. Indeed, the forces are expected to be roughly proportional to the number of doubly bound complexes, which is itself roughly proportional to the length of the overlap (fig. 3, $S_1$). So if the asters are pushed apart from the equilibrium position, or if microtubules shorten, attractive force will rise and repulsing forces will fall, leading the asters to move back together. The reverse is true, if the asters get too close, or if microtubules grow. Although the forces are also proportional to the number of complexes in solution, the equilibrium is in fact quite insensitive to the concentration of motors (www). This is because both attractive and repulsive forces are produced by the same complexes, only in different configurations. The balance of forces in 2D and 3D is essentially the same as described in 1D, but there are additional requirements. For example, to sustain equilibrium, the motors need to bend the microtubules which extend off-axis, and continuously bring them to overlap with microtubules from the other aster. The microtubule rigidity should also be able to sustain their longitudinal load. The simulations show that all the necessary requirements are indeed met.

To test the uniqueness and accessibility of the equilibrium position, the simulations leading to $S_1$ were repeated, keeping the same parameters, but starting from different initial configurations and using different random number sequences. The average reliability was high (9 reruns failed out of 227), and many individual $S_1$ solutions had very high reliability. For example, figure 5 shows a simulation in which 100 repetitions never failed to produce the same interaction. All this suggests that $S_1$ determines a unique equilibrium position which can be reached from any connected initial configuration.

**Screen 3b:** The screens described so far considered motors that would detach immediately when they reach the ends of their track (high $P_{ends}$). However, in reality a motor can halt for a short interval when it reaches the end of a microtubule. Therefore, the hetero-complex screen 3a was repeated, this time allowing the motors to stay at the ends of growing microtubules for variable periods of time ($P_{end} = P_{end\_growing}$ varied between 0.04 and 50 $s^{-1}$), but not at the ends of shrinking microtubules ($P_{end\_shrinking}$ is fast). Three new classes of interactions were found, with anti-parallel microtubule overlaps: $S_2$, $S_3$, $S_4$ (fig. 3). The first one, $S_2$, which represented ~3 % of the simulations, is particularly interesting, because it is the only one which led to a *partial anti-parallel microtubule overlap* (fig. 3).

In $S_2$, the balance of forces is between motors which have reached the end of a microtubule, and motors which did not, both acting on anti-parallel overlaps (fig. 3). Hence, the pushing forces are produced by a population of motors whose number is roughly proportional to the length of microtubule overlap, and pulling forces are produced by a population of motors which is proportional to the number of microtubule ends, i.e. more or less constant. Because pushing forces decrease directly with aster-aster distance, while pulling forces do not, the asters will be dragged closer or further away, until they find a distance at which equilibrium is reached and this equilibrium will be stable.

A plot shows the requirements on the motor speeds (fig. 6). $S_2$ only covers part of the domain $u \cdot v < 0$ and $u + v > 0$. Motors slower than ~0.25 µm/s seem to be unable to achieve $S_2$ equilibrium. Simulations with static microtubules produced $S_2$ with speeds over the entire domain (data not shown). Therefore, the apparent additional constraint is related to the microtubule growth and/or shrinkage ($V_{growth}$= 0.16 µm/s, $V_{shrink}$= -0.25 µm/s). Indeed, a pre-requisite for $S_2$ is that motor complexes actually reach the ends of growing microtubules. This is only possible if the plus-end directed motor is faster than the growth speed of the filaments ($u > V_{growth}$). Furthermore, to be able to pull a microtubule by its plus end towards the center of the other aster (fig. 3, $S_2$), the minus end directed motor should also be faster than the growth speed ( $|v| > V_{growth}$). Because these conditions apply on the effective speed at which motors move, which is usually only a fraction of $V_{max}$, the slowest speed found for $S_2$ is not $V_{grow}$, but a somewhat higher value ~0.25 µm/s. Further screening should produce a limit closer to $V_{grow}$.

**Screen 3c:** A hetero-motor complex screen in which the motors could stay attached even to shrinking microtubules was performed ($P_{end} = P_{end\_growing} = P_{end\_shrinking}$ between 0.04 $s^{-1}$ and 50 $s^{-1}$). With this, a plus end directed motor who has reached the plus-end of a shrinking microtubule would remain attached, following the end towards the center of the aster. The solutions $S_2$ are now over all the region $u \cdot v < 0$ and $u + v > 0$ (fig. 6), meaning that the additional constraints on the motor speeds were released by allowing them to hold on shrinking microtubule ends.

The difference between $S_1$ and $S_2$ is revealed by looking directly at the forces: in $S_2$, motor-complexes attached to the region of anti-parallel microtubule overlap are pushing, while in $S_1$, they are pulling (fig. 3). $S_3$ and $S_4$ are quite similar to $S_1$, and a portion of the forces in $S_3$ or $S_4$ are always due to $S_1$ contributions. They are not described here in detail, but can be understood from figure 3, and examples of simulations can be found online (www). Last, we examined what is the distance between the asters in the solutions $S_1$ and $S_2$. Figure 3 shows that in one dimension, with static microtubules, the center to center distance in the *full anti-parallel overlap* $S_1$ is mostly determined by the length of the microtubules. Simulations with dynamic microtubules are more complex, but can be compared because they all include microtubules following the same dynamic, which have on average the same length of ~7 µm. The examples in figure 2, top have aster-aster distance of 7.4 +/- 1 µm, 8.2 +/- 0.24 µm, 13.9 +/- 1.1 µm, respectively from left to right (average and standard deviations from 100 simulations), although they all realize the same equilibrium $S_1$. Hence the characteristics of the motors and microtubules contribute together to the equilibrium. Overall averages show that asters in a *partial anti-parallel overlap* $S_2$ are 11.4 +/- 4.2 µm apart, while they are distant by 9.3 +/- 3.9 µm in a *full anti-parallel* *overlap* $S_1$. Asters are further apart in $S_2$ than in $S_1$, but the two distributions largely overlap; consequently with dynamic microtubules, $S_1$ and $S_2$ are not as easily distinguishable from the aster to aster distance as figure 3 might suggest.

## Discussion

Recent progress in light microscope technology has revealed the dynamic nature of biological organization. Many structures in the cell are in fact self-organized steady state assemblies resulting from continuous stochastic interactions at the molecular level. In order to describe and understand these structures in quantitative terms, there is a need for new tools allowing the analysis of the collective behavior of molecules. The mitotic spindle is a good example of a self-organizing cellular structure, whose function is to segregate chromosomes into the newly forming daughter cells. Spindle morphologies can vary considerably amongst different cell types; for example, some spindles have well focused poles, while others do not, or some pull on the chromosomes using an anti-parallel microtubule overlap to counterbalance the forces, while others do so by attaching to the cell cortex. The spindle is host to many interesting phenomenon, e.g. the dynamic instability of microtubules, the local stabilization of microtubules by chromatin and by MAPs, or the action of various molecular motors. Integrating all these to explain the different spindle morphologies will most certainly require the use of computer simulations. Here I have modestly started to integrate some simple properties into a simulation and I used it to study a single feature known to occur in spindles: that microtubules originating from opposite poles can form stable anti-parallel overlaps. As a starting point, one can do this ignoring many other aspects of spindle assembly by artificially reducing the spindle to two asters of microtubules, with which the feature is first tested. It is still unknown if motors alone can produce stable anti-parallel interactions. This question was here addressed theoretically, ignoring for example that force production by microtubule assembly/disassembly might be necessary to achieve this feature.

The systematic computer screening shows that hetero-complexes of plus- and minus-end directed motors can produce stable interaction patterns between microtubule asters. These figures resemble spindles because they have similar features: two poles and overlapping anti-parallel microtubules between them. On the contrary, the various mixtures of homo-complexes computed only led to fusion or full separation of the two asters. The failure to find stable anti-parallel overlaps with homo-complexes could be attributed to limitations of the screen, e.g. the number of simulations was too small, but there might be more fundamental reasons. First, homo-complexes acting on just two microtubules do not produce anti-parallel overlap but instead organize them into a parallel configuration (fig. 7). Second, in the simple situation studied, a partial overlap in which both pulling and pushing forces are proportional to the amount of microtubule overlap is bound to be unstable.

The screens were computed in one and two dimensions with similar results (not shown), and the solutions also exist in three dimensions (www). The solutions $S_{1-4}$ could uphold static (not shown) or dynamic microtubules, with some differences in their requirements on the motors, as was shown for $S_2$. Because the simulations are based on discrete stochastic events, the solutions can also tolerate some variation in the parameters that define these events: e.g. binding, unbinding rates and number of motors (www). The different solutions show us simple ways in which a balance of forces can reliably be achieved (fig. 3). Importantly, the simulations showed that all these situations are stable: if the microtubules disassemble, or if by chance the asters are displaced too far, the forces will be modified, and the asters will be dragged back together. The reverse is true when the microtubules grow, or if the asters get too close. The mechanical stability of the equilibrium is a simple but essential property realized in each solution. Surprisingly, the simulations suggest that the solutions are *unique* stable equilibria, which can be reached from any initial configuration, and this makes them quite interesting from a cell biological point of view.

The simulation was initially developed to study an in-vitro mixture of pure proteins, and its first validation came from comparison with experiments (Surrey et al. 2001, Nédélec and Surrey, 2001). The model for the motors is inspired from the measured behavior of kinesin, which is a processive motor whose speed varies almost linearly with the load (Hunt et al., 1994, Svoboda and Block, 1994). This is not intended to be realistic in every detail, but could represent other processive motors, or regulated processive assemblies of non-processive motors. However, a limitation is that the model might not correspond to isolated non-processive motors.

Inside the cell, many effects could allow more elaborate types of equilibria. For example, microtubule can produce forces without motors (Holy et al, 1997), and some motors might influence microtubule stability (Hunter and Wordeman, 2000). Mechanisms of biological regulation or the presence of chromosomes could change the composition or properties of some molecules in time or space, or as a feedback of the present organization. Some key elements of the spindle, such as the proposed 'matrix' (Kapoor and Mitchisson, 2001) might still be missing. Clearly, cells have many other ways to realize a stable balance of force. Yet, the solutions described in this work are simple and emerge from core properties of microtubules and motors. They represent conceptual solutions that could be generated in-vivo by various molecular mechanisms.

The solutions points to two interesting possibilities which can be tested. The first one is that a pause of the motors at microtubule ends gives them additional morphogenetic properties, confirming previous proposals (Hyman and Karsenti, 1996; Kurse and Julicher, 2000; Nédélec and Surrey, 2001). The second and new possibility is that hetero-complexes with both plus- and minus- end directed motor activities have the capacity to produce anti-parallel overlaps, from which stable interactions can be made. Such complexes could easily be built in the cell. This might not necessarily be by direct and stable interactions between different motor proteins, but rather by transient association, allowing its regulation both in space and in time during mitosis. There is some evidence for interaction between motors of opposite directionality (Blangy, 1997), but, to my knowledge, direct physical association between motors of opposite polarity *in vivo* has not been proven so far. It is worth pointing out that, in the spindle of some organisms, there is a microtubule flux towards the poles (Waterman-Storer and Salmon, 1997). In the model, this flux would be integrated in first approximation by subtracting the flux speed from the speed of all motors. The putative plus- and minus-end directed motor complexes could be built from the association of a static microtubule binding protein with a plus-end directed motor, which would need to be faster than the microtubule flux. In this situation, one can even imagine that two plus-end directed motors of very different speeds would make an effective plus-minus complex.

In summary, computer simulations have been described and used to systematically explore how a stable balance of forces can be achieved between two asters and soluble motor complexes. Four solutions were found (Fig. 3), which achieve stable equilibrium by combining simple properties of polar filaments and directed motors. The different qualitative requirements on the motors for each of these solutions were expressed simply, and should be testable experimentally. All the solutions rely ultimately on a common principle: by their movement, the hetero-complexes place the motors on the opposite side from where they would naturally go: a minus-end directed motor near the plus ends of a microtubule, or vice versa. With this simple rule, we can already imagine other solutions. For example any mechanism which leads to the localization of a plus-end directed motor to the center of asters, or a minus-end directed one near the plus ends of microtubules, while leaving the motor free to grab passing microtubules can also lead to stable interactions between asters.

## Methods

The principle of the simulation, simulated dynamics, is intuitive: from an initial configuration, the future of the sample is calculated in small successive time steps. All the forces and movements of individual filaments and motors are calculated by solving the equations of motion, which are set according to the laws of classical mechanics. These motions include simple and natural processes: The filaments diffuse, grow and shrink, and respond elastically to deformation. The motors diffuse, can bind and move onto the filaments, exerting forces on them, and eventually unbind. All interactions are based on "first principles". All parameters could in principle be directly obtained from single molecule experiments. Importantly, this makes the conditions that we find on parameter values simple to interpret. Because the thermal noise present in the world of molecules is essential to their function, the simulation is largely stochastic. The probabilities of most events (e.g. binding and release of motors) are calibrated according to rates given as parameters to the program, but their exact timing is not predictable. The simulations thus include at least some of the noise present in nature.

All features of the natural world can not, and should not, be retained. Among the many simplifications introduced, the biggest is that two filaments will only feel their respective presence if they are somehow linked by motors. As a consequence, the filaments can cross or overlap freely without any steric or hydrodynamic interactions. In some of the simulations that we discarded (fusion and oscillatory solutions), the distance between the asters come close to zero. The solutions $S_{1-4}$ on the contrary were selected because the distance between the asters is always greater than 2 μm, and steric interaction should not alter them significantly. With the methods described, a PC build in 2001 could each day simulate a fully connected structure of a thousand microtubules. Such a system is roughly comparable in size to an animal mitotic spindle. Hence, the limitations are not so much numerical, but really in the intrinsic biological complexity of the subject.

For each time step, the objects in the system are considered in random order, in each of the following: **A:** Dynamic instability of microtubules: stochastic transitions between growing/shrinking state; lengthen/shorten according to state. **B:** Solve the motion of the filaments, considering their elasticity, the action of motor complexes, and Brownian forces. **C:** Stochastic attachment of free motors, detachments of bound ones, optional detachment if the force is above a threshold, and displacement along the filament. **D:** Diffusion of free motor complexes. These processes are detailed below.

**Microtubules:** Simulated microtubules are linear, infinitely thin, oriented objects which mechanically behave like inextensible elastic rods (Feynman, 1989). The position of a microtubule is represented by N+1 points $M_i$ for i = 0 to N, with all the distances $|M_i M_{i+1}|$ being equal. $M_0$ is the minus-end and $M_N$ is the plus end. The length L of the microtubule can take any value, and as it grows or shrinks, points are dynamically added or removed to ensure that at all times, N achieves the minimum of the absolute value of (R - L/N), while staying greater or equal to one. R is a parameter of the model called the filament section length. Hence, all segments $|M_i M_{i+1}|$ on the same microtubule have the same length L/N, which is always close to R, but L/N can vary from microtubule to microtubule.

Smaller values of R ensure more accurate computation, but require more processor time. One consideration when choosing R is that it should be smaller than the radius of curvature of the filaments, and this can be checked after the simulation has been performed. The force needed to fold a filament on a radius of curvature R is $\sim E / R^2$, where E is its mechanical bending modulus. Knowing the motor's maximum forces, this formula provides a guideline to initially choose R. The simulations presented here were calculated with R = 1.2 μm ($E/R^2 \sim 14$ pN), and occasionally R = 0.5 μm.

Simulated microtubules are not stretchable, but can bend elastically under external forces or Brownian motion. For any three consecutive points $M_{i-1}$, $M_i$ and $M_{i+1}$, the program calculates $F = E \cdot (N/L)^3 \cdot (M_{i+1} - 2M_i + M_{i-1})$, where E is the bending modulus (20 pN.μm² for microtubules, Kurachi et al. 1995) and L/N the distance between consecutive points. The force 2F is applied to $M_i$, while -F is applied to $M_{i-1}$ and $M_{i+1}$. This linear elastic torque realizes the theoretical value (Feynman, 1989) under small deformations.

**Asters:** An aster is a set of microtubules attached at their minus end with static and permanent Hookean links. Its structural integrity is independent of the activity of the motors. The microtubules are also attached laterally to their side neighbors, at some distance from the center (0.75 μm). In three dimensions the aster is built similarly, by using a triangulation of the isocahedron as a template. The resulting structure is always a very rigid tensegrity construction, in which the microtubules are regularly distributed. Asters move or rotate as a whole without much stretch (typically below 2 nanometers) in their static links.

**Dynamic instability:** The microtubule's dynamic instability is modeled according to experimental data obtained for centrosome nucleated asters in mitotic Xenopus egg extract (Dogterom et al., 1996). The minus-end of the microtubule, in the center of the aster, is static. The plus-end is either shrinking or growing, and the transitions between these two states *depend* only on the microtubule length L (Dogterom et al., 1996), making longer microtubules less stable: catastrophe frequency ($s^{-1}$) is equal to $L \cdot 0.003$, and rescue frequency ($s^{-1}$) is $(13 - L) \cdot 0.00333$ (when L is in μm). The growth and shrinkage speeds are constant, +10 and -15 μm/min respectively. For simplicity, microtubules do not shrink below 1 μm, hence their number is constant.

The microtubules have, under these conditions, a mean length of ~7 micro-meters, with a (large) standard deviation of ~3 μm. They spend ~54 % of their time growing, ~31 % of their time shrinking, and ~15 % of their time at their minimal length 1 μm, waiting for a rescue.

Simpler models where the transition probabilities are independent of the microtubule length can also be used. In fact, the simulations were initially performed with static microtubules of fixed length and produced very similar results to those presented here (data not shown). Dynamic instability is here considered, because sudden catastrophes and fast microtubule disassembly makes the asters very variable and hence unpredictable structures. Consequently, the simulations demonstrate better that the solutions are very stable with respect to the number of microtubules and their lengths.

**Motors:** The minimal processive entity that can bind to a microtubule, and move on its surface is called a *hand*. A hand is always associated to a second one, forming a bi-functional soluble *motor complex*.

**Hands:** Each hand (a single motor) is either detached, or attached to a microtubule, and transitions between these two states occur in the following way: A free hand is at a position defined by the complex it belongs to. From this position, it may bind, with given rate $P_{on}$ (s$^{-1}$), on every microtubule geometrically closer than a given parameter ε (μm), the reach of the motor. When ε is small, binding is limited by diffusion (Nédélec et al., 2001). Bound hands are entirely characterized by a record of the microtubule they are attached to, and the position (or abscissa) at which they are bound, counted from the minus end. Bound motors detach stochastically with a rate $P_{off}$ (s$^{-1}$), but otherwise move along the microtubule in a direction defined by the intrinsic directionality of the motor. A bound motor which has, by its movement, reached the end of a microtubule detaches with a different unbinding rate $P_{end}$ (s$^{-1}$). We can imagine that $P_{off}$ is a property of a moving motor, which goes through rounds of ATP hydrolysis, while $P_{end}$ is a property of an immobile motor probably arrested in one configuration at the end of its track. Both rates can therefore have unrelated values.

**Stochasticity:** All the discrete events in the simulation, binding and unbinding events, catastrophes and rescues are modeled stochastically. To decide if a possible event with a rate P (s$^{-1}$) should be performed or not during a period dt (10$^{-2}$ s), the program draws a pseudo random number x between 0 and 1 (Matsumo and Nishimura, 1998). The event is performed if x < P·dt. Practically, this procedure limits the rates to P < 50 s$^{-1}$ (0.5 / dt). Rates above 100 s$^{-1}$ are equivalent and results in an immediate and thus non-stochastic action.

**Unbinding from the microtubule ends:** The detachment of a motor from the plus end of a microtubule could be different whether this end is shrinking or growing. In all generality there are two rates: $P_{end\_shrinking}$ and $P_{end\_growing}$. For clarity, the exploration is limited to three situations: (1) Screens 1, 2 and 3a: Motors always detach at all ends ($P_{end\_shrinking}$ and $P_{end\_growing}$ both greater than 100 s$^{-1}$, i.e. immediate). (2) Screen 3b: Motors detach fast from shrinking ends and slowly from growing ends ($P_{end\_shrinking}$ immediate and $P_{end} = P_{end\_growing}$ randomly taken within 0.04 and 50 s$^{-1}$). (3) Screen 3c: Motors detach equally from shrinking and growing ends ($P_{end} = P_{end\_shrinking} = P_{end\_growing}$ randomly taken within 0.04 and 50 s$^{-1}$). The minus-ends of the microtubules are not static, and minus-end directed motors always unbind from this end with the same rate $P_{end}$.

**Motor movement:** All interactions between motors while bound to the same microtubule are neglected. Motors pass or cross each other without interacting. Similarly, the microtubule lattice never saturates, and more motors can always bind. This simplification should be true at low density of motors, but might break down at high densities. Simulated microtubules are 'smooth' and their structure is neglected, they also have similar properties all along their length, and are all equivalent. Motors always bind to the geometrically closest point on the microtubule, and not to specific sites that would be defined by the tubulin lattice. Similarly, if a motor moves at speed V, its abscissa will be simply increased during a time dt by V·dt, whether or not this represents a multiple of the tubulin lattice. An unloaded motor is moving at a speed given by a parameter $V_{max}$, plus-end directed if $V_{max} > 0$ or minus-end directed if $V_{max} < 0$. We will see later how the effective speed V of the motor is determined by the possible load applied to the motor, created when the motor is part of a complex linking two microtubules.

**Motor complexes:** A motor complex is a set of two hands linked by a Hooke's spring (i.e. a linear force). These two hands behave independently, with the exception that a complex cannot be bound twice to the same or to successive segments on the same microtubule. The states of its two hands determine the state of the complex, and its behavior. Free complexes diffuse with a coefficient D (20 $\mu m^2/s$). Complexes attached to only one microtubule are transported along this microtubule, at the unloaded speed of the attached hand, without exerting force. On the contrary, a complex attached to two microtubules effectively links them with a force F=K·dx, which is proportional to the separation vector between the two hands dx. For simplicity, the stiffness K (pN. $\mu m^{-1}$) is the same for all complexes.

**Force-velocity relationship:** Four similar models for the force-velocity curves of hands were used, which are defined by two parameters: a maximum force $F_{max}$ (always positive), and a maximum speed $V_{max}$. Movement occurs at decreasing speed for increasing load; the simplest model implements a full linear force-velocity dependency $V = V_{max} (1-F_{axis}/F_{max})$, where the scalar $F_{axis}$ is the projection of the force F on the axis of the microtubule, taking into account the directionality of the motor: $F_{axis} > 0$ for a force which resists the natural movement of the motor, and $F_{axis} < 0$ for a force that helps it. With this simple model, a motor pulling a passive load, for example a bead, would never exceed its maximum force $F_{max}$, or its maximum velocity $V_{max}$. However, this motor can be forced backward, for example if it is pulled by other motors.

In experiments, a motor like kinesin has never been observed to make a backward step. The model therefore includes two mechanisms to prevent backward motions. The first limits the linear range to $-F_{max} < F_{axis} < F_{max}$. Outside this range, the motor is either moving at $2V_{max}$, or held immobile, respectively. The second mechanism imposes detachment on motors if they satisfy $F^2 > dim\ F_{max}^2$. The test is here on the norm of the vector force F, hence the factor dim (the number of dimensions in space), to allow a motor to effectively reach a stall force along the filament axis, with an additional off-axis component. Three models for the motors are derived from the full linear one by including one or both of these restricting mechanisms. However, apart from oscillations, there is no strong qualitative difference between all the four models in the present study. The oscillations required that the detachment of a motor depends on its load, which is only achieved here in the models which include detachment at maximum force ($F^2 > \dim F_{max}^2$).

**Elasticity of the motors:** We can generally expect that molecules act as non Hookean springs with a non-zero resting length. For simplicity the simulated complexes follow a Hooke's law with a null resting length (kinesin stiffness seems to be constant, around 400 pN/µm; Kawaguchi and Ishiwata, 2001). This contributes greatly to the simplicity and to the numerical stability of the final algorithm, but implies that the stiffness K, the stalling force $F_{max}$ and the reach ε depend on each other:
1. The equipartition theorem of statistical mechanics states that at a temperature T, any spring stores on average an energy $\dim/2k_BT$ (this is also true in the simulations). If the stiffness of the spring is K, this corresponds to a random force of magnitude $<F^2>=\dim \cdot K \cdot k_BT$. For the motors which detach at their maximum force, this random force should be smaller than $F_{max}$. Therefore to simulate kinesin ($F_{max} = 5$ pN), K as high as 500 pN/µm is appropriate, but to simulate a weaker motor with a stalling force of 0.5 pN, the stiffness needs to be lower, for example K = 5 pN/µm.
2. Consistency of the model also requires that the force at the time of binding be inferior to the maximum force of the motor, i.e. $K \cdot \varepsilon < F_{max}$.
For the screen presented here, K and $F_{max}$ were drawn randomly between 5 and 60 pN/µm, and 0.5 to 2 pN, respectively. To satisfy (1) with a safety factor of 3, only sets which satisfied $\dim \cdot K \cdot k_BT < 1/9 F_{max}^2$ were kept. To satisfy (2), ε was always equal to $F_{max}/K$, and altogether varied between 25 and 200 nm.

**Linearity of the forces:** A position X on a microtubule between two consecutive points $M_i$ and $M_{i+1}$ is calculated as $X = M_i + (M_{i+1}-M_i) |M_iX|/|M_iM_{i+1}|$. The force exerted by a complex is a difference of two such positions $F = K(Y - X)$. F is acting on the first microtubule at the position X, and -F applies to the second microtubule at point Y. For the first microtubule, F is further split, according to the position of X between $M_i$ and $M_{i+1}$: $F_i = F |XM_{i+1}| / |M_iM_{i+1}|$ applies in $M_i$, and $F_{i+1} = F |M_iX| / |M_iM_{i+1}|$ applies in $M_{i+1}$. Hence the contribution of a motor link is linear.

**Initialization and confinement inside a box:** Motors are initially free and uniformly distributed over the simulation box of 60 x 60 µm. Confinement is achieved by reflecting boundaries. To ensure that the asters are close enough to interact, they are set randomly within the 12 x 12 µm central square. The microtubules are all initially 1 µm long, in a growing state, but loose this synchrony after ~100 seconds. Confining the filaments was not necessary, as the box is much larger than their size.

**Movement of the filaments:** As usual on the scales of micrometers, the movement is dominated by viscosity, and inertia is neglected (Berg, 1993). The speed of each microtubule-point is proportional to the sum of the forces acting on it: $dM_i/dt = mob_i.F_i$. The mobility $mob_i$ of the N+1 points in a microtubule of length L is the same for all i: $mob_i = H(N+1)( 4\pi \text{ viscosity } L )^{-1}$. For simplicity, no correction is made for the filament orientation, or for hydrodynamic interactions. To yield more accurate drag forces, the factor $H = \log_e( 2 \text{ µm} / 25 \text{ nm} ) = 4.38$ corrects for the tubular shape of the microtubule (Hunt et al., 1994). It is based on a hydrodynamic cut-off length of 2 µm, and on the diameter of the microtubules, 25 nm. All the simulations are performed with a viscosity of 0.05 pN.s. $\mu m^{-2}$, or ~50 times the viscosity of water, and are intended to represent the conditions inside a cell or a developing egg, of which the viscosity could be as high as 100 times the viscosity of water (Hiramoto, 1970).

**Integration of the motion:** Altogether, the model reduces to the points defining the microtubule's positions, connected by forces which are linear in the coordinates of these points (elasticity, motor complexes and aster static links). The problem is technically difficult because the total number of variables is large (typically 2,000 for the current work). The first order differential equation is solved implicitly with a constant time step dt = $10^{-2}$ s. Practically, the coordinates of the points $M_i$ of all microtubules at time t are pooled in a vector $M^t$. Scanning through the interactions builds a matrix A such that $A.M^t$ is the force acting at time t on the points. To represents the Brownian forces in the system, a random vector E is calibrated to yield $<E_i^2> = k_BT / ( mob_{avg} . dt )$, with $mob_{avg} = H (4\pi \text{ viscosity } R)^{-1}$, and $k_BT$ for a temperature of 37°C. The Jacobian projection P defined by the conservation of the lengths $M_iM_{i+1}$ for each microtubule is calculated. Finally, the implicit system $M^{t+dt} - M^t = dt \text{ mob} \cdot P( A.M^{t+dt} + E )$ is solved to get $M^{t+dt}$, with mob the diagonal of all the $mob_i$. The constraints are not perfectly preserved, and the points are further moved, to restore the length of each microtubule. This is only a small contribution to the movement of the points, usually below 0.1 nm per time step and per point (the diffusion is ~40 nm/step). This correction is done while conserving the barycenter of each microtubule, to avoid introducing any systematic error. The most intensive part of the computation is usually to solve the linear system, and processor times vary mostly with the size of the matrix, i.e. the number of microtubules considered. The simulations took here on average ~20 minutes in 1D, ~3 h. in 2D, and ~5 h. in 3D, on a 700 MHz Pentium III Linux PC.

**Internal controls:** To check that the choice of time step dt, section length R, and size of the box did not affect the outcome of the simulations beyond numerical precision, simulations were repeated with all possible combinations of R = 0.5 or 1 μm, dt = 5 or 10 ms within a simulation box of 60 or 120 μm, while conserving the concentration of the motor complexes. All these different choices produced virtually identical outcomes (www). Indeed, the numerical precision achieved by the standard choice of R, dt, and box size is sufficient with respect to the Brownian forces in the model.

**Parameter values:** Appropriate values for kinesin would be roughly $F_{max}$ ~ 5 pN, $V_{max}$ ~ 0.8 μm/s, K ~ 100 pN/μm, ε ~ 50 nm, $P_{on}$ ~ 50 /s, $P_{off}$ ~ 1 /s. These estimates are used as guidelines to set the ranges, in which the values of the motor parameters are drawn randomly for any individual simulation (see table 1). The number of motors in the simulation is also chosen randomly, their force-velocity relation is any one of the four described. The number of microtubules in the aster is chosen between 20 and 80, as this is typically the number of microtubules nucleated by a centrosome in Xenopus egg extracts. In the 1000 seconds simulated, very slow motors would not have enough time to produce a stable situation between the asters. In the automatic screen, these slowly evolving interactions would be mistakenly selected, because they persist during the simulated period. To limit this, I *a-priori* did not consider simulations in which the speeds of the motors are below 0.01 μm/s, or for which rates are below 0.04 /s.

**Acknowledgements:** I thank Eric Karsenti and Thomas Surrey for their precious input in this work, their daily support and our discussions. I also thank Stephanie Blandin, Damian Brunner, Marileen Dogterom, Michael Knopp, Stan Leibler, Oleg Lerner, A. C. Maggs, Iain Mattaj and Jean-Claude Nédélec.

## Figure Legends

**Figure 1: Summary of the screens.** Screen 1 was performed with two kinds of homo-complexes, with all sorts of configurations of minus- or plus- end directed motors. It produced fusion of the asters, or their full separation. Screen 2 is inspired by the putative configuration of the biological motors involved in the spindle. Screen 3a was performed with one kind of hetero-complex. It produced fusion, full separation, oscillations, and one type of non-fused stable interaction, solution $S_1$. To achieve $S_1$, the speeds of the motors u and v need to satisfy $u+v < 0$ and $u \cdot v < 0$, as depicted in the diagram. Screen 3b is a variation in which the motors could hold on the microtubule ends. Four solutions are found, as discussed in the text.

**Figure 2: Examples of stable interactions between dynamic asters**, with hetero-complexes. Top: solutions of type $S_1$. The speeds (μm/s), of the two motor forming the complex are: left: 0.35, -0.91; middle: 0.31, -0.73; right 0.6, -0.83. Bottom: example of solutions $S_2$, speeds are: left: 0.95, -0.45; middle: 0.47, -0.45; right 0.89, -0.64. Below each example is plotted the distance between the two asters (μm) as a function of time (seconds). It is not possible to distinguish from these views the different solutions. See animations on (www).

**Figure 3: The balance of forces in the solutions.** Schematic asters in one dimension only have two opposing microtubules; radiating from a common center represented by a black diamond. All solutions are built from one kind of hetero-complex, with two speed u and v, which must satisfy the conditions specified here. Solution $S_1$ is found even when motors immediately detach from the end of the microtubules, while the others are obtained when the motors can stay at the end (in this situation, u or v is replaced by e). Pushing or pulling is schematically represented here by the tilt of the complex, which is a consequence of the relative movement of both motors. The attractive or repulsive nature of the forces can be deducted by mentally trying to restore the complexes in a vertical position.

**Figure 4: Symmetry arguments.** (A) Two asters can have anti-parallel overlaps, but also parallel ones when they are close. (B) On an anti-parallel overlap, hetero complexes of speeds u and v produce attractive force if $u + v < 0$, or repulsive force if $u + v > 0$. Both possible motor configurations produce forces in the same direction. (C) On a parallel overlap, if $u \neq v$, the two configurations result in opposite forces. If they are equally probable, these forces cancel each other. A homo-complex (u=v), does not produce any force on parallel microtubules, and stabilizes the overlap.

**Figure 5: Reliability of a solution $S_1$**: Figures produced by 49 simulations, all performed with the same parameter set, but with different initial configurations and random number sequences. The figures are all similar, showing the reliability in which the parameter set determines the evolution of the system towards a unique interaction configuration. Each picture covers 30x30 μm.

**Figure 6: Probing the constraints on the speeds.** Each symbol represents one simulation, and is plotted here as a function of the unloaded speeds $V_{max}$ of the two motors in the complex. Dots: $S_1$ produced in screen 3a. Circles: $S_2$ produced in screen 3b, in which the motor could stay attached only to growing microtubules. Pluses: $S_2$ produced in screen 3c, when the motor could stay attached also to shrinking ends.

**Figure 7: The zipper effect:** Schematically, the action of homo-complexes on two microtubules produces parallel microtubule overlap, while hetero-complexes produce anti-parallel overlaps.

**Table 1: Parameters of the simulation**

| Parameter description | Symbol | Value or range of values |
|---|---|---|
| **Model-specific parameters:** | | |
| Time step | dt | $10^{-2}$ s. |
| Filament section | R | 1.2 µm |
| Simulation box | | square, 60x60 µm |
| **Fixed parameters:** | | |
| Total simulated time | | 1000 s. |
| Temperature | $k_BT$ | $4.2\ 10^{-3}$ pN µm |
| Fluid viscosity | viscosity | 0.05 pN s µm$^{-2}$ |
| Microtubule's bending modulus | E | 20 pN µm$^2$ |
| Microtubule dynamic instability: catastrophe and rescue frequencies | | rescue in s$^{-1}$: ( 13 - L ) * 0.0033<br>catastrophy in s$^{-1}$: L * 0.003<br>L : length of microtubule in µm. |
| Growth and shrinking speeds | | +10 and -15 µm/s |
| Complexes diffusion | D | 20 µm$^2$ /s |
| **Independently varied parameters:** | | |
| Number of microtubules per aster | | between 20 and 80, same for both asters. |
| motor links rigidity | K | between 5 and 60 pN/µm |
| Number of complexes:<br>1 parameter in the screens 3a,b,c<br>2 parameters in the screens 1 and 2. | | between 500 and 15,000 |
| Motor's speed ( 2 parameters) | $V_{max}$ | between -1 µm/s and +1 µm/s, $V_{max}$ is rounded to a multiple of 0.01, and $V_{max} = 0$ is excluded. |
| Motor's stalling force (2) | $F_{max}$ | between 0.5 pN and 2 pN,<br>with $F_{max}^2 > 9.\text{dim}.K.k_BT$ |
| Motor's binding rate (2) | $P_{on}$ | between 0.04 s$^{-1}$ and 50 s$^{-1}$ |
| Motor's unbinding rate (2) | $P_{off}$ | between 0.04 s$^{-1}$ and 50 s$^{-1}$ |
| Motor's unbinding rate from microtubule end | $P_{end}$ | $P_{end}$>100 s$^{-1}$ in screens 1, 2, 3a;<br>0.04 s$^{-1}$ to 50 s$^{-1}$ in screens 3b,c. |
| Motor's Force-velocity type (2) | | one of four models (see methods) |
| **Derived parameters:** | | |
| Motor's reach distance | ε | = $F_{max}$ / K (see methods) |

Figure 1

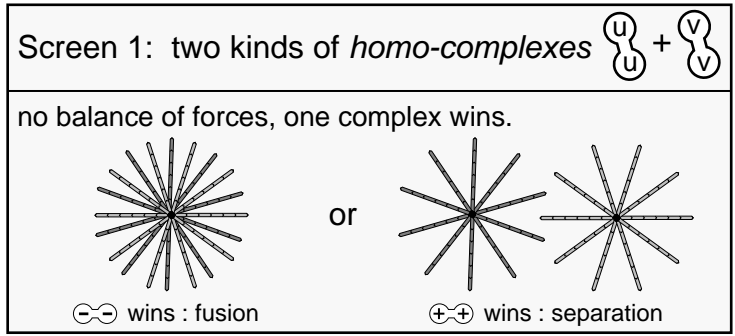
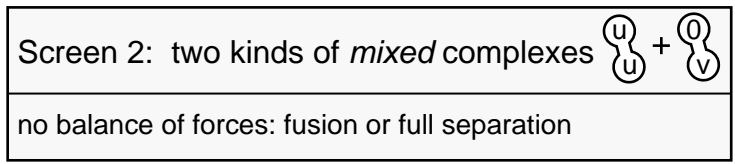
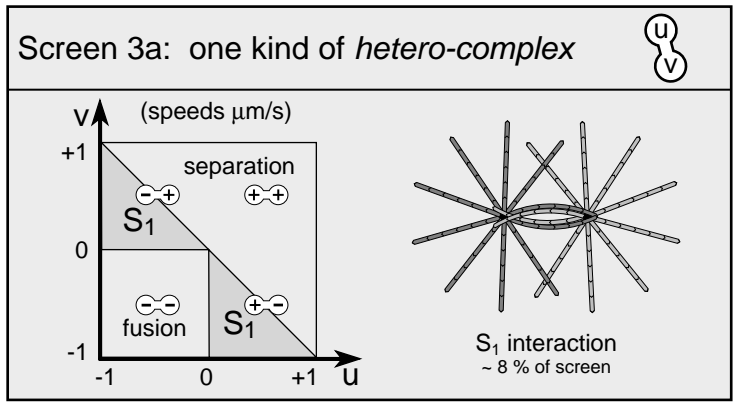
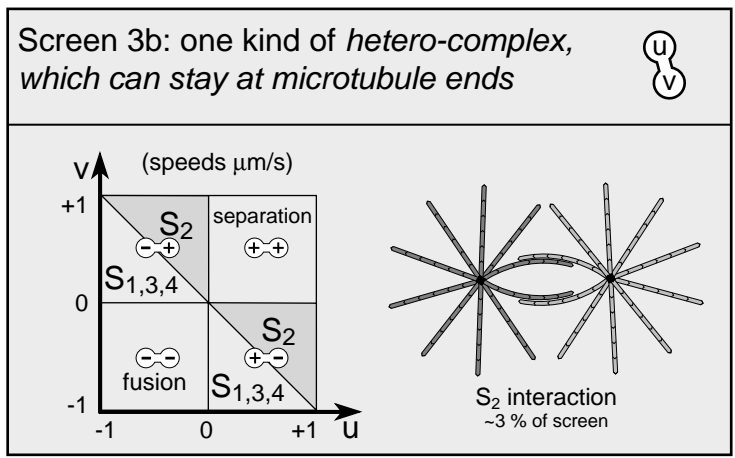

Figure 2

S₁

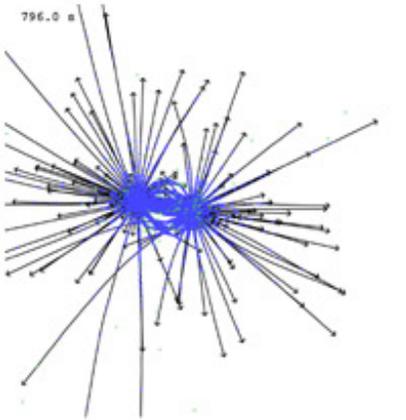
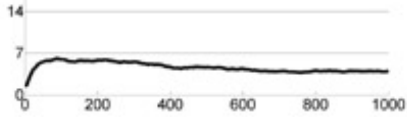
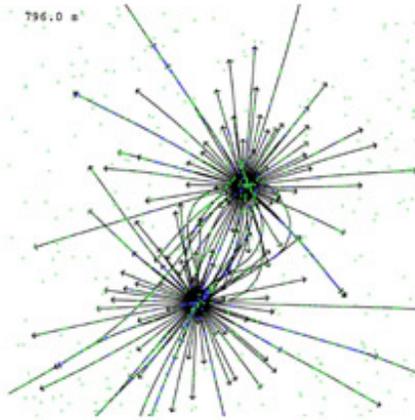
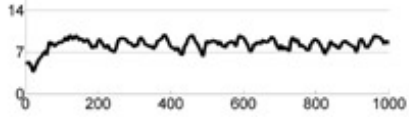
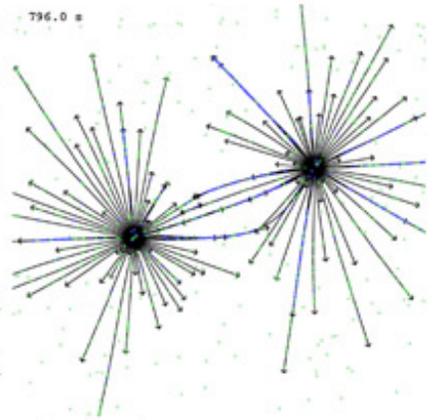
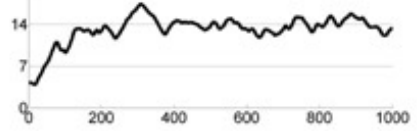

S₂

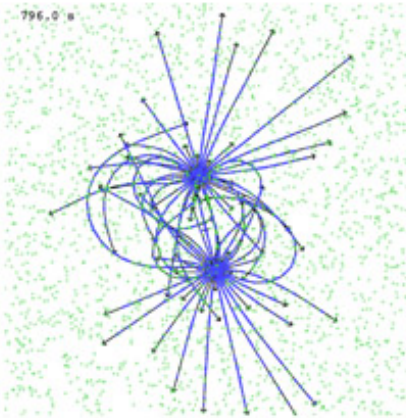
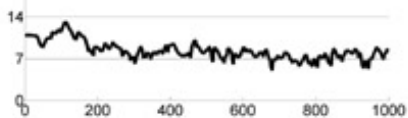
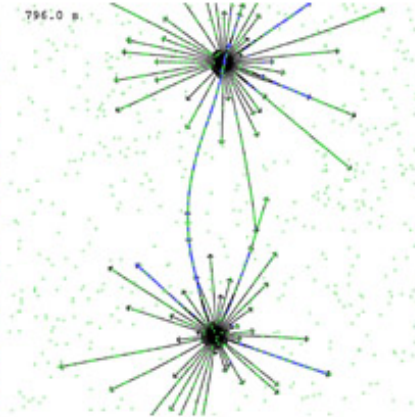
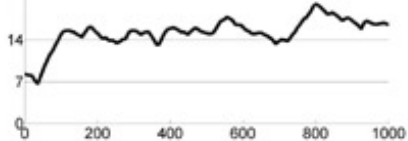
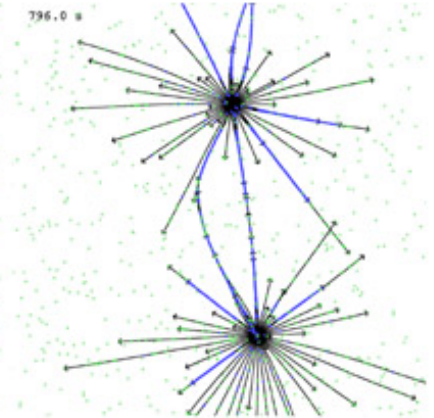
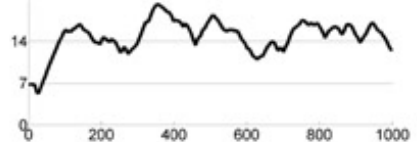

Figure 3

## Solutions where motors detach immediately at microtubule ends

$S_1$
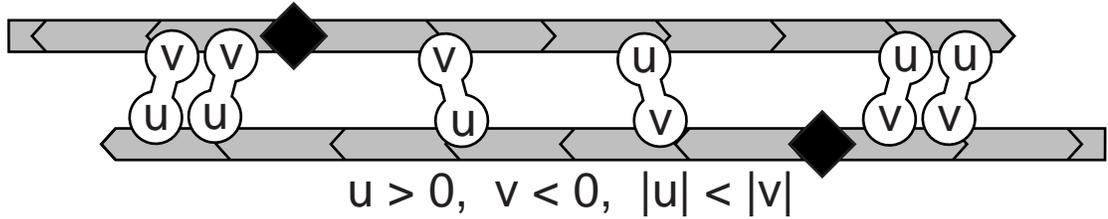
$u > 0, \; v < 0, \; |u| < |v|$

## Solutions where motors can halt at microtubule ends

$S_2$
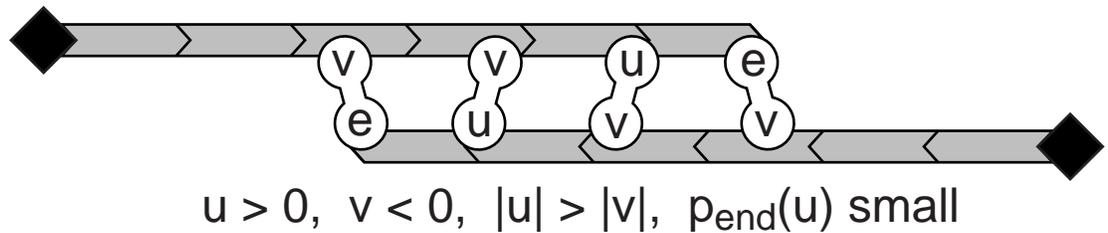
$u > 0, \; v < 0, \; |u| > |v|, \; p_{end}(u) \text{ small}$

$S_3$
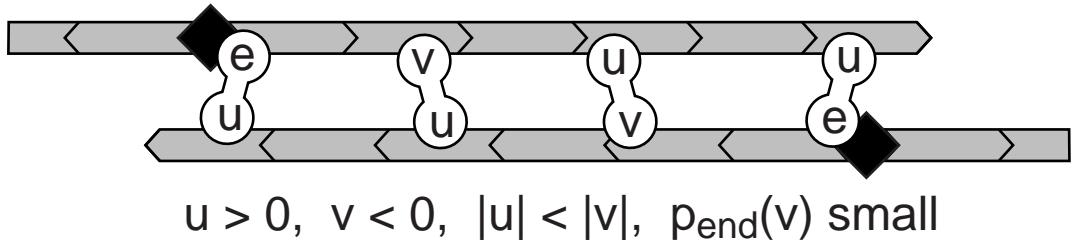
$u > 0, \; v < 0, \; |u| < |v|, \; p_{end}(v) \text{ small}$

$S_4$
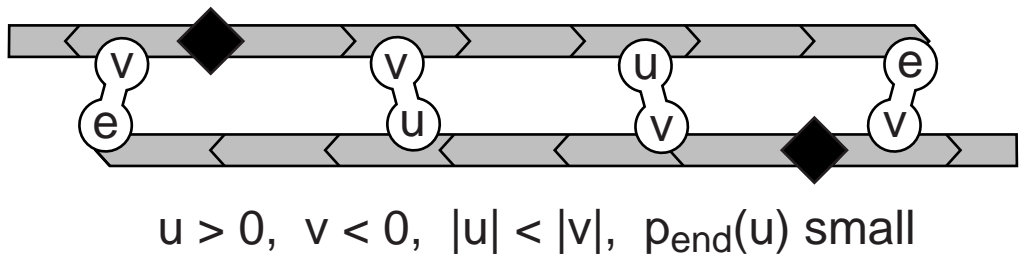
$u > 0, \; v < 0, \; |u| < |v|, \; p_{end}(u) \text{ small}$

Figure 4

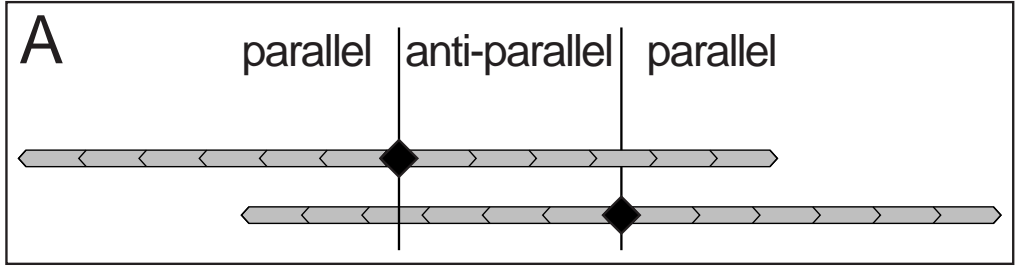
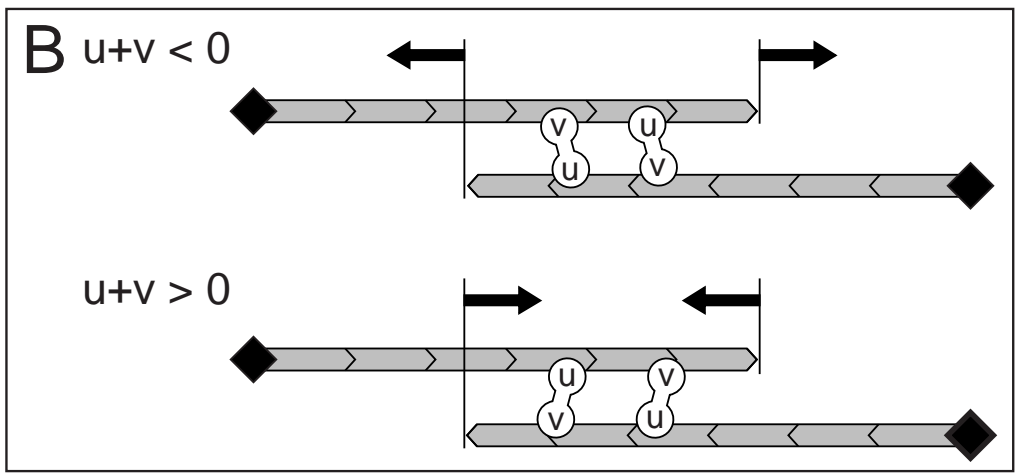
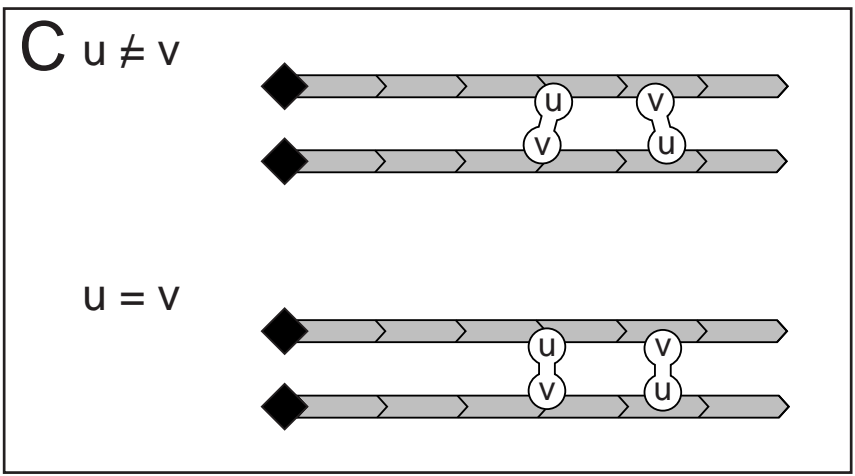

Figure 5

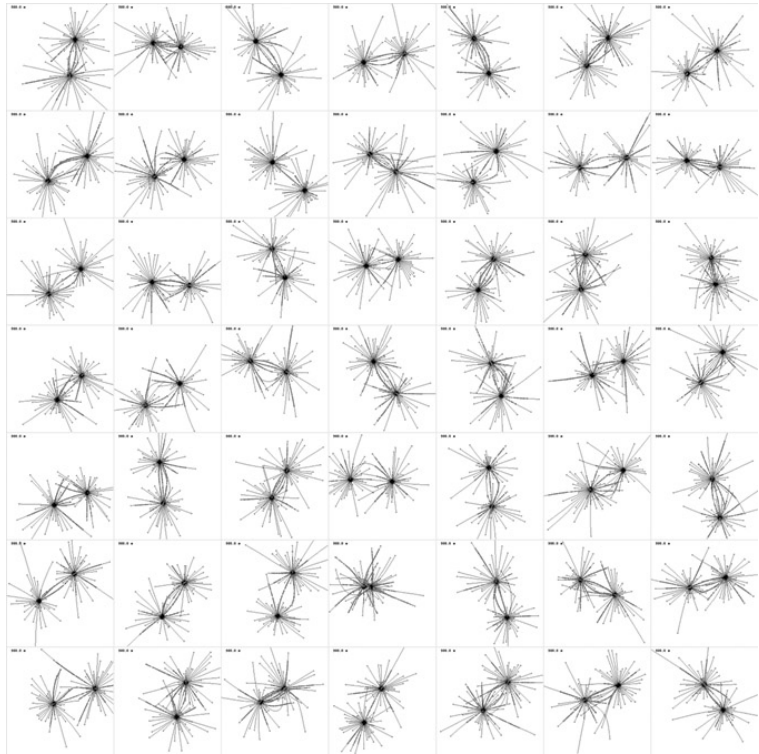

Figure 6

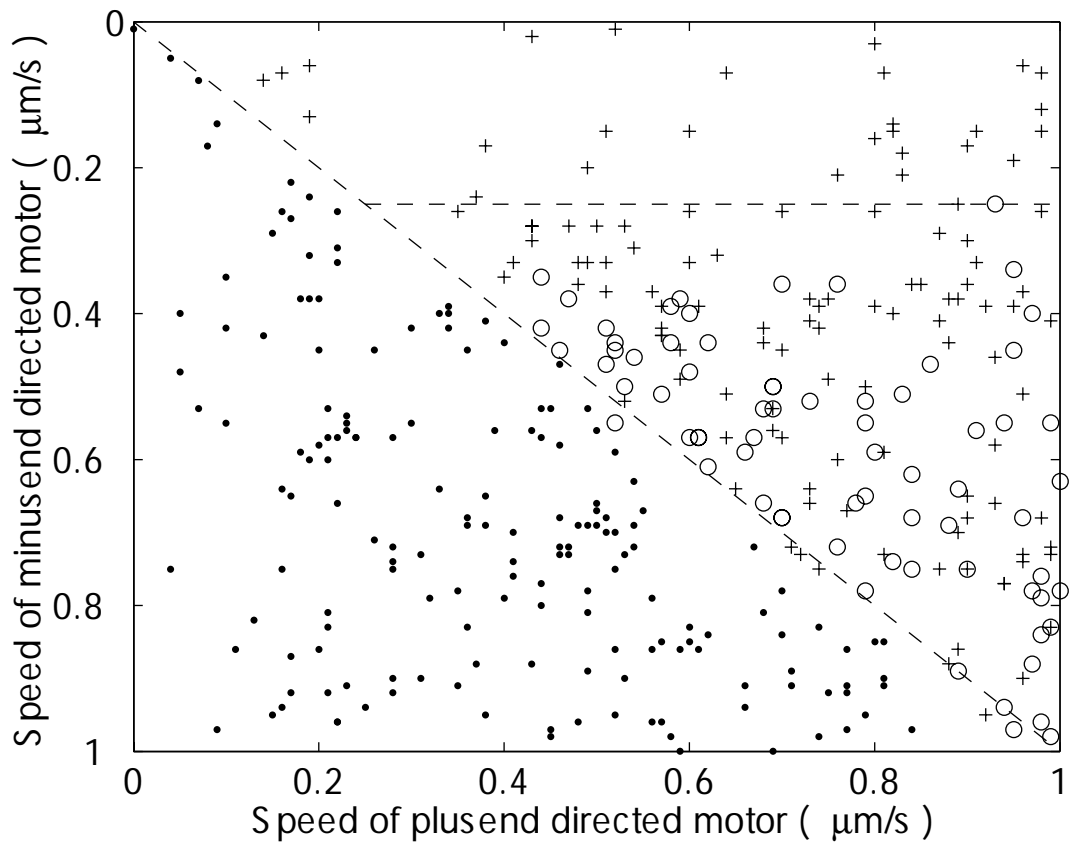

Figure 7

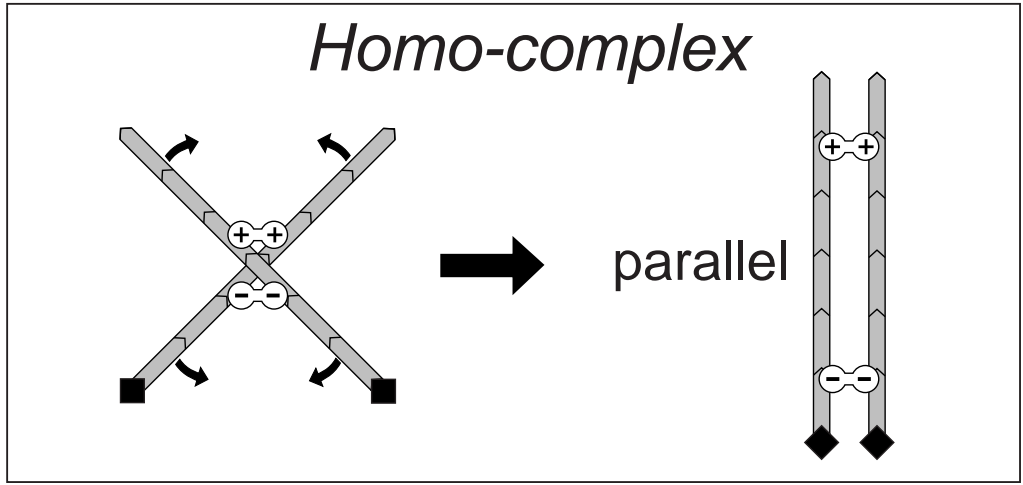

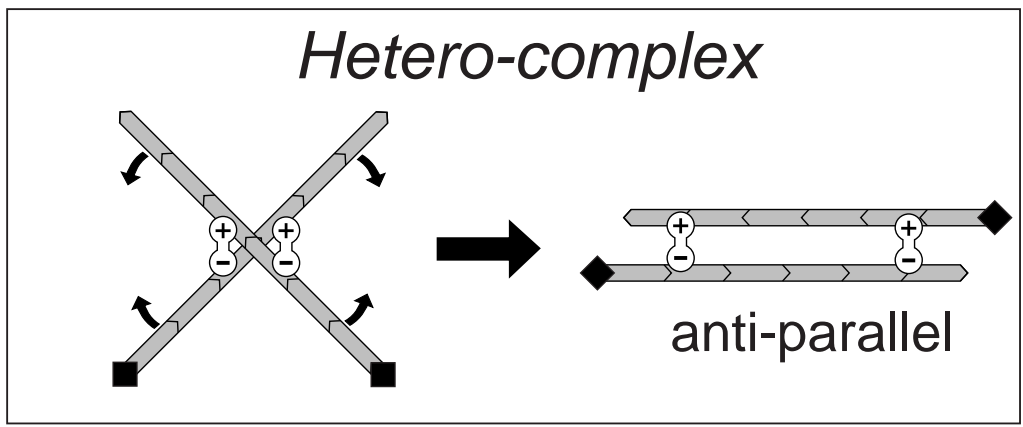